# Ultra-Narrow Linewidth Brillouin Lasers with Nanokelvin Thermometry

WILLIAM LOH,* SIVA YEGNANARAYANAN, FREDERICK O'DONNELL, AND PAUL W. JUODAWLKIS

*Massachusetts Institute of Technology Lincoln Laboratory, Lexington, MA 02421, USA*
*William.loh@ll.mit.edu

**Abstract:** Ultrastable lasers serve as the backbone for some of the most advanced scientific experiments today and enable the ability to perform atomic spectroscopy and laser interferometry at the highest levels of precision possible. With the recent and increasing interest in applying these systems outside of the laboratory, it remains an open question as how to realize a laser source that can reach the extraordinary levels of narrow linewidth required and yet still remain sufficiently compact and portable for field use. Critical to the development of this ideal laser source is the necessity for the laser to be insensitive to both short- and long-term fluctuations in temperature, which ultimately broaden the laser linewidth and cause drift in the laser's center frequency. We show here that the use of a large mode-volume optical resonator, which acts to suppress the resonator's fast thermal fluctuations, together with the stimulated Brillouin scattering (SBS) optical nonlinearity presents a powerful combination that enables the ability to lase with an ultra-narrow linewidth of 20 Hz. To address the laser's long-term temperature drift, we apply the narrow Brillouin line as a metrological tool that precisely senses a minute change in the resonator's temperature at the level of 85 nK. The precision afforded by this temperature measurement enables new possibilities for the stabilization of resonators against environmental perturbation.

## 1. Introduction

The ultra-low expansion (ULE) cavity-stabilized laser [1−4] is presently the prominent technology for realizing laser linewidths of 1 Hz or below. This level of spectral purity is required for a variety of applications spanning basic and applied science such as trapped-ion quantum computing [5], precision spectroscopy [6], optical-atomic clocks [7, 8], gravitational wave interferometry [9], optical frequency division [10], and frequency metrology. Currently, no other laser technology exists that provides both the ultra-narrow linewidth and long-term temperature stability of the ULE cavity-stabilized laser, as the propagation of light in any material other than vacuum adds both loss and sensitivity to temperature drift. Despite the advantages afforded by the ULE cavity, two significant drawbacks hold the laser back from being a universal solution for all applications requiring a stable laser source. These drawbacks are the bulky size of the ULE cavity laser (~ 1 $m^3$) and the laser's susceptibility to vibration noise, both of which have rendered the laser difficult to use outside of a laboratory environment [11, 12].

We show here that optical gain provided by the stimulated Brillouin scattering (SBS) optical nonlinearity [13], when combined with a high quality factor ($Q > 10^8$) resonator, presents a way to achieve a stable laser source that addresses the limitations of size and vibration sensitivity currently constraining the ULE-cavity design. The recent advances of optical frequency combs generated through four-wave mixing [14, 15] in a nonlinear microresonator and the development of narrow-linewidth lasers created through optical injection locking [16] or through stimulated Brillouin [17−20] and Raman scattering [21, 22] all serve to highlight the promise offered by high-Q resonators [18, 23−28] as precise frequency references [29]. In particular, the potential of the SBS laser technique results from its inherent ability to suppress the noise of the pump source [30]. This linewidth reduction in conjunction with techniques that stabilize the pump laser to the cavity resonance [31], effectively provides two stages of linewidth reduction that can bring the SBS linewidth below that of other laser technologies. For serving as a precise frequency reference, the central challenge facing most microresonator devices is their inherent sensitivity to temperature fluctuations, which causes the lasing frequency to fluctuate at time scales longer than the thermal response time of the cavity [32]. In this work, we demonstrate a SBS laser that fundamentally addresses this core issue and reaches an integrated linewidth of 20 Hz by utilizing a high-Q resonator comprising 2-meters of polarization-maintaining (PM) fiber. The resonator's large mode volume critically suppresses thermo-refractive fluctuations [33] and increases the laser's resistance in response to temperature change. In distinct contrast to previous reports of SBS in optical fiber resonators [34], our demonstration also uses a low coupling ratio (~5 %) into the resonator to achieve operation near critical coupling. As we will show later, this distinction significantly increases the quality factor of the cavity and results in over two orders of magnitude improvement in noise for our SBS laser.

The frequency wander in lasers is an equally challenging problem separate from a laser's linewidth that results from the slow temperature drift of the laser's environment. In silica glass, for example, a 1°-C shift in temperature results in a 1.65-GHz shift of the lasing frequency. This frequency shift is 8 orders of magnitude larger than the 20-Hz linewidth of our laser, and therefore the laser's temperature must be stabilized to ~10 nK to keep the SBS laser center frequency within one linewidth of its original position. The necessary precision in temperature stabilization represents the major challenge all lasers face when the laser cavity consists of any material other than vacuum. We address this issue by introducing a novel self-referenced scheme that senses the temperature drift of the cavity through the exceptional frequency precision afforded by our laser. Our temperature sensor combines the differential temperature sensitivity of the cavity's two orthogonal polarization modes [29. 35−37] with the exquisitely narrow SBS lasing line to detect minute temperature fluctuations as small as 85 nK.

## 2. Results

### 2.1 SBS laser configuration and characteristics

We generate the SBS laser light by sending the output of an integrated planar external-cavity laser pump through a high-Q optical-fiber resonator. The resonator consists of a tunable coupler with two of its ends spliced together to form a fiber ring 2 meters in length. The corresponding cavity free spectral range is 100 MHz, which for a Brillouin shift of 10.9 GHz means that 108 modes are skipped between the pump and SBS lasing resonance. However, since the Brillouin gain bandwidth is ~50 MHz, only a single longitudinal mode reaches oscillation despite the long cavity length.

Figure 1(a) shows a diagram of our experiment configuration comprising a pair of independent SBS lasers and a frequency-noise characterization system. The frequency noise is measured from the detected microwave beat signal from the pair of SBS lasers. In each SBS laser, a pump laser is first transmitted through a phase modulator and then through a semiconductor optical amplifier to boost the output power up to ~15 mW. The pump signal is next directed through a circulator and into the optical resonator for the generation of the SBS light. The remainder of the pump light that bypasses the resonator is collected on a photodetector and then is subsequently used in a Pound-Drever-Hall [31] (PDH) scheme to stabilize the pump frequency to the cavity resonance. Meanwhile, the reverse-propagating SBS laser signal is directed back through the circulator and combined with the other cavity's SBS signal to generate a 3.7 GHz microwave beat after photodetection.

Figure 1(b) depicts resonator mode scans corresponding to three coupling ratios of the resonator. The coupling ratio is controlled by the transverse separation of two polished optical fibers mounted in contact. Critical coupling occurs at a coupling ratio of 1.3% with an associated resonance linewidth of 420 kHz, which implies that our fiber resonator reaches an intrinsic Q of 920 million. In order to maximize the output power of our SBS laser, we instead choose to operate at a coupling ratio of 5.6% where the loaded cavity Q reduces to 170 million but permits more power to be coupled out of the cavity. Figure 1(c) shows the SBS output power as a function of the pump power input to the resonator coupler. The SBS laser threshold power is 4.6 mW and the slope efficiency is 79% but rolls off at higher pump power (~20 mW) when a second SBS line begins to reach oscillation threshold. For the laser noise measurements described below, we operate with 11-mW pump power and generate 5-mW SBS output power, which corresponds to a very high optical-to-optical conversion efficiency of 45%. We note that the coupling ratios used here represent a marked decrease compared to the ~50% coupling previously used for fiber SBS lasers [33]. At 50% coupling, the cavity Q not only becomes artificially degraded (> 10×) compared to operation at critical coupling but also the vast majority of input pump power bypasses the resonator and becomes unusable for SBS generation. Often, the resonator length is necessarily increased by over an order of magnitude to compensate for the degradation in Q, which in turn results in system instability [38] due to multiple oscillating modes and increased sensitivity to environmental perturbations.

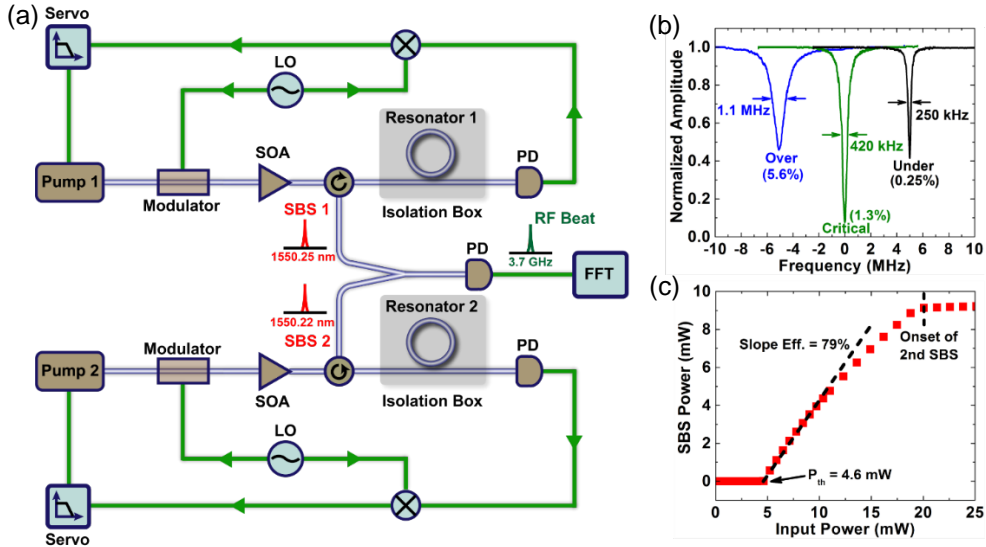

**Fig. 1.** SBS laser system setup and operation. (a) Schematic of a pair of identical SBS lasers and laser characterization systems. Each SBS laser comprises a pump laser, a phase modulator, a semiconductor optical amplifier (SOA), a fiber resonator, and a photodiode (PD). The pump is locked to the cavity resonance through demodulation via a local oscillator (LO). (b) Laser scans over the mode profile of the resonator for varying coupling ratios. Critical coupling occurs at 1.3% coupling corresponding to a resonance width of 420 kHz. (c) SBS laser output power as a function of the supplied input optical pump power. The threshold power is 4.6 mW, and the slope efficiency is 79% until the second SBS Stokes oscillation occurs at 20 mW input.

One of the greatest assets gained by using a large mode-volume resonator is the long time constant for the system to settle into thermal equilibrium. The resulting insensitivity to thermal fluctuation is essential not only to isolate the SBS laser from disturbances in the outside environment but also to prevent the internal coupling of intensity noise to changes in the resonator temperature. Previously, these thermal fluctuations were identified as the dominant limitation in SBS laser performance [32], which degraded the SBS noise by as much as 4 orders of magnitude at low frequencies. For our fiber resonator, Fig. 2(a) shows the measured thermal response of the SBS laser frequency when the resonator temperature is changed by both varying the input pump power and varying directly the heat applied to a section of the resonator. For the case of direct heater modulation, the response of the SBS laser continues to roll off even for frequencies down to 100 mHz signifying a thermal time constant longer than 1 second. When the optical power is instead modulated to indirectly change the resonator temperature, the same roll off occurs but comes to a halt beyond 0.4 Hz. At this point, the resonator frequency shift reaches the Kerr nonlinearity limit where the material index change becomes dominated by a shift in optical intensity rather than a shift in temperature.

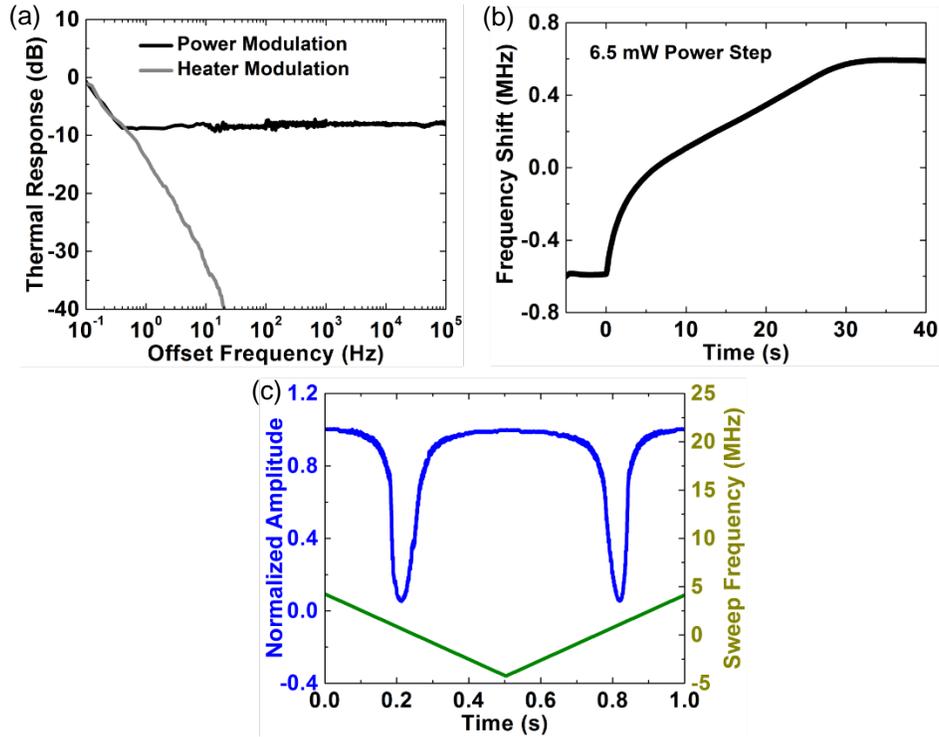

**Fig. 2.** SBS laser thermal response. (a) SBS laser thermal response corresponding to the cases of input optical power modulation and heater modulation. Both cases show a continuous roll off in response starting at 100 mHz, but for optical power modulation, the Kerr limit is reached after 0.4 Hz. (b) Time domain response of the SBS laser frequency for a 6.5 mW step increase in input pump power. The measured rise time is 24.6 s. (c) Pump laser sweep over a resonance mode showing no visible effects of mode asymmetry from thermal bistability. The scan speed is intentionally made slow to allow the pump to spend 70 ms within the cavity linewidth.

Since the roll off is still present at the lowest frequency of 100 mHz, the thermal response was instead determined in the time domain by measuring the time it takes for the SBS laser frequency to respond to an input optical power shift of 6.5 mW [Fig. 2(b)]. Ideally, this response should be exponential, and we believe the slight deviation from a pure exponential is due to slight environmental temperature shifts over the 40 second measurement. With a 10%-90% rise time of 24.6 s, we ascertain the thermal time constant to be 11.2 s, which we use to normalize the 3-dB point of the frequency response in Fig. 2(a). The benefit of the long time response on the resonator's temperature stability is clearly observed in the resonator mode scan of Fig. 2(c). In Fig. 2(c), an 8.3-MHz scan of the input pump frequency from high to low and vice versa occurs in 1 second, which equates to a time of 70 milliseconds spent within the resonance linewidth. Because the time on resonance is much lower than the cavity's thermal response time, no effects of thermal bistability are observed, and the modes from the forward and backward scans are seen to be symmetric. In contrast, the modes of typical microresonator systems exhibit significant asymmetry due to the thermal broadening that occurs when the resonance responds to the frequency shifts of the pump laser [39]. For these faster-responding microresonator systems, the pump frequency noise is transferred over to the cavity resonance.

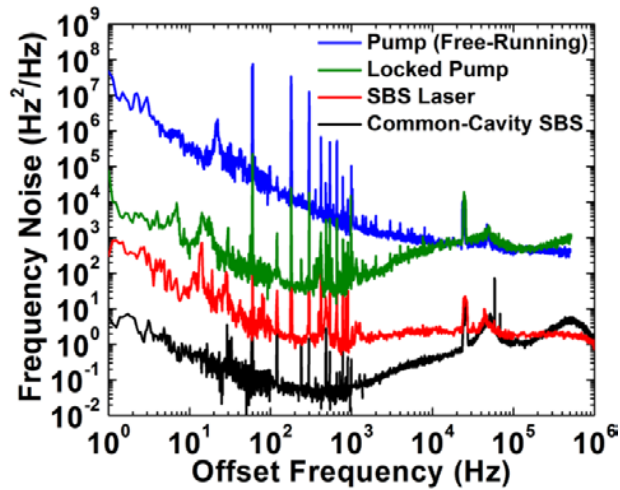

**Fig. 3.** Laser frequency noise comparison. Measurements of frequency noise corresponding to the pump laser (blue), the PDH-locked pump (green), and the SBS laser (red). The SBS laser receives an additional factor of 30 dB noise suppression from the locked pump case, but this suppression diminishes below 30 kHz offset frequency. By measuring the noise using two orthogonal polarization SBS lasers generated from one common resonator, the common-mode noise is suppressed, and the SBS noise reaches a new lower floor (black). This floor is now 30 dB reduced from the locked pump case at all offset frequencies.

The excellent thermal stability of our resonator makes it ideal for use as a narrow-linewidth SBS laser source. Figure 3 shows the frequency noise measured for the pump laser, the PDH-locked pump, and the SBS laser. In free-running operation, the pump laser exhibits an integrated linewidth [40] of 3 kHz and reaches a frequency noise level of $3\times10^5$ $Hz^2/Hz$ at 10 Hz offset frequency. By locking the pump laser to the cavity resonance, the noise within the locking bandwidth is improved to a value of 830 $Hz^2/Hz$ at 10 Hz offset, which yields an integrated linewidth of 270 Hz. We note that although a 3 kHz linewidth pump laser was employed here, with sufficient servo gain, the PDH locking stabilizes the pump to a linewidth of ~270 Hz even when significantly broader linewidth pump lasers are used. Therefore, the linewidth of the SBS laser does not strongly depend on the availability of a narrow linewidth pump source. Finally, by achieving SBS lasing in the resonator, the noise receives one more factor of improvement due to the intrinsic SBS suppression of the pump noise [30]. This noise suppression occurs in addition to the improvement gained from pump locking and enables the SBS frequency noise to reach 30 $Hz^2/Hz$ at 10 Hz offset frequency, which corresponds to an integrated linewidth of 20 Hz.

For our SBS laser, the ideal value of noise suppression [30] is 30 dB as is observed when comparing the locked pump noise to the SBS noise above 30 kHz offset. However, below 30 kHz, we observe this suppression to decrease as the SBS noise flattens and increases again below 100 Hz offset. The 2 Hz$^2$/Hz white noise limit of the SBS noise is a result of intensity noise in the SBS laser coupling to frequency noise through the nonlinear Kerr effect. To determine the ideal performance of our SBS laser, we bypass this limit by comparing two orthogonal polarization SBS lasers generated from a single common cavity. This configuration suppresses the common-mode intensity noise as well as any mechanical and vibration noise present, and the resulting SBS frequency noise is shown in Fig. 3 (black line). The measured noise decreases to 0.5 Hz$^2$/Hz at 10 Hz offset frequency and yields a significantly reduced integrated linewidth of 2.6 Hz, which shows the ultimate potential for a Hertz-class SBS laser given that the residual common-mode noise is mitigated. From Fig. 3, the 30 dB ideal SBS suppression over the locked pump is also now apparent at all frequencies with the noise increasing past 1 kHz due to the 30 kHz PDH servo bandwidth.

*2.2 SBS laser temperature sensing and stabilization*

In addition to the SBS laser's short-term linewidth performance, the long-term frequency drift of the laser is a separate issue that is also equally important to address. For laser systems where the optical intensity resides in any material other than vacuum, a small temperature change results in a massive change in the cavity's resonance frequency (e.g., 1 °C induces ~1.65 GHz shift for silica glass). In order to both detect and correct for this frequency drift, we employ the dual- polarization temperature sensing technique recently developed for microresonator systems [29, 35−37], which utilizes the birefringence of the orthogonal polarizations to create a differential measurement of temperature change. We further improve the resolution of this technique by using the exquisitely narrow linewidth of the SBS laser to detect the resulting difference in cavity resonance shifts induced by temperature. Figure 4(a) shows a schematic of our SBS temperature sensor consisting of two independent pump lasers that are phase modulated and amplified before they are combined onto one common path. Prior to combining the two lasers, one path is rotated to the orthogonal polarization, and then the two pump lasers are sent through a circulator into a single fiber resonator. Each pump laser probes a separate mode belonging to the two orthogonal polarizations and is locked on resonance using the PDH technique. Afterwards, the two generated orthogonal polarization SBS laser outputs traverse backwards through the circulator and are separated using a polarization beam splitter. One path is rotated to once again match the polarization of the other path, and the SBS outputs are combined on a photodetector to measure their microwave beat.

A change in frequency of the microwave beat directly corresponds to a change in the temperature seen by the optical mode of the resonator. Figure 4(b) illustrates this point where at room temperature (ΔT = 0 °C), the resonator modes aligned to the two orthogonal fiber polarizations are initially 29 MHz apart with the fast axis mode trailing the slow axis mode. When a 0.6 °C temperature shift is applied (as measured by a thermistor), the relative position of the two modes changes such that the two modes are now coincident in frequency. Finally, with a further total applied temperature shift of 1.2 °C, the fast axis mode moves to a position 29 MHz ahead of the slow axis mode. By locking the two orthogonal polarization pumps each to a separate mode, the difference in frequency of the pump lasers directly reflects a shift in the resonator's temperature. Moreover, as the SBS output is generated at a frequency offset from the pump, the information of the temperature shift is also imprinted on the SBS microwave beat. Figure 4(c) shows the efficiency of using this dual-mode technique for temperature sensing where for every 1 MHz change in the resonator mode's absolute frequency, the relative separation of the modes changes by 30 kHz.

The resolution of the dual-mode temperature sensor is ultimately set by the linewidth of the lasers used for interrogating the mode separation. This limitation arises because the frequency noise on the lasers transfers over to the frequency of the resulting microwave beat and hence gives the appearance of a change in temperature when no such change exists. It is important to note that when the microwave beat is then used as the feedback signal to servo the resonator's temperature, this noise becomes imprinted on the resonator as a correction to a temperature change that never occurred. For these reasons, we define the sensitivity of our dual-mode Brillouin temperature sensor using the SBS laser's linewidth rather than by using the noise of the already-stabilized in loop microwave beat as was traditionally done in dual-mode sensor demonstrations [35–37]. Figure 5(a) shows the measured lineshape of the dual-polarization beat between two SBS lasers and two pump lasers with the beat-note linewidth converted to a temperature resolution via Fig. 4(c). Owing to the SBS laser's intrinsically narrow linewidth that critically becomes even lower here (integrated linewidth = 2.6 Hz) with the common-mode noise suppression of Fig. 3, our Brillouin sensor (dn/dT = $11.6 \times 10^{-6}$) [41] is able to detect temperature changes with 85 nK resolution as seen in Fig. 5(a). In contrast, our measurements using the locked pump laser (integrated linewidth = 270 Hz) in place of the SBS, which corresponds to the traditional dual-mode sensor configuration, yielded a temperature sensitivity of 12.9 µK.

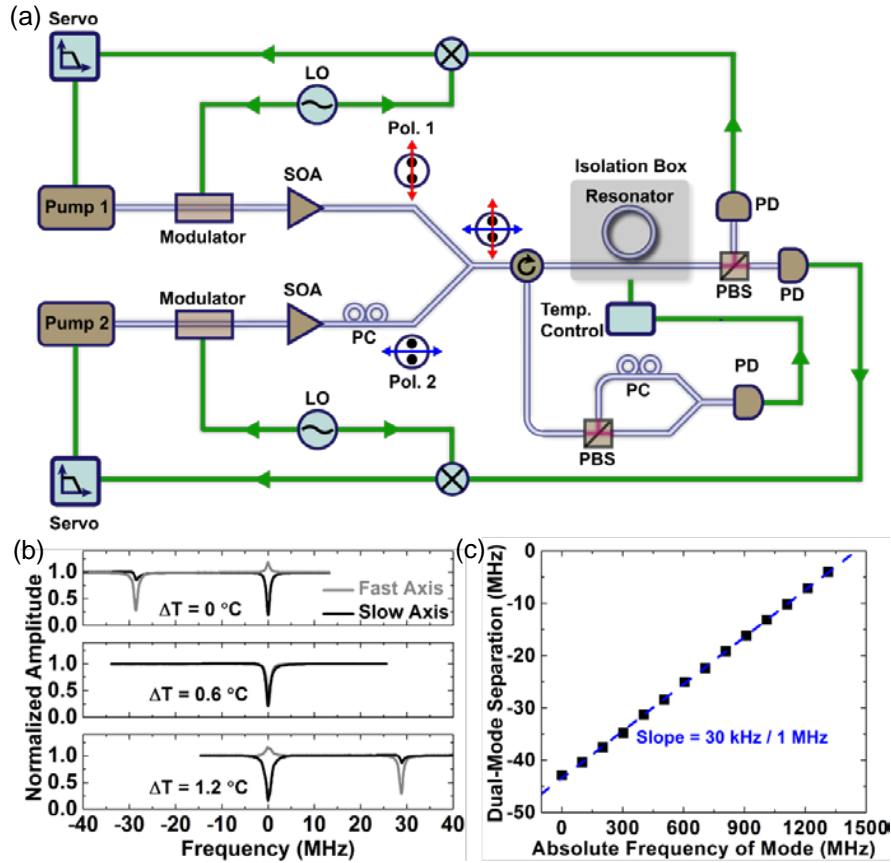

**Fig. 4.** SBS laser dual-mode thermometry. (a) Configuration of the dual-mode SBS laser comprising 2 pump lasers that are phase modulated, amplified, and photodetected for locking to two orthogonal-polarization modes of a single resonator. One of the pump lasers is rotated using a polarization controller (PC) to the orthogonal polarization, and the two polarizations are separated using a polarization beam splitter (PBS). (b) Measurement of the shift in orthogonal polarization modes with temperature. A 1.2 °C temperature change corresponds to a ~58 MHz change in mode separation. (c) Efficiency of the dual-mode separation in response to temperature relative to the response of a single resonator mode. The dual-mode separation changes by 30 kHz for every 1 MHz shift in the cavity resonance.

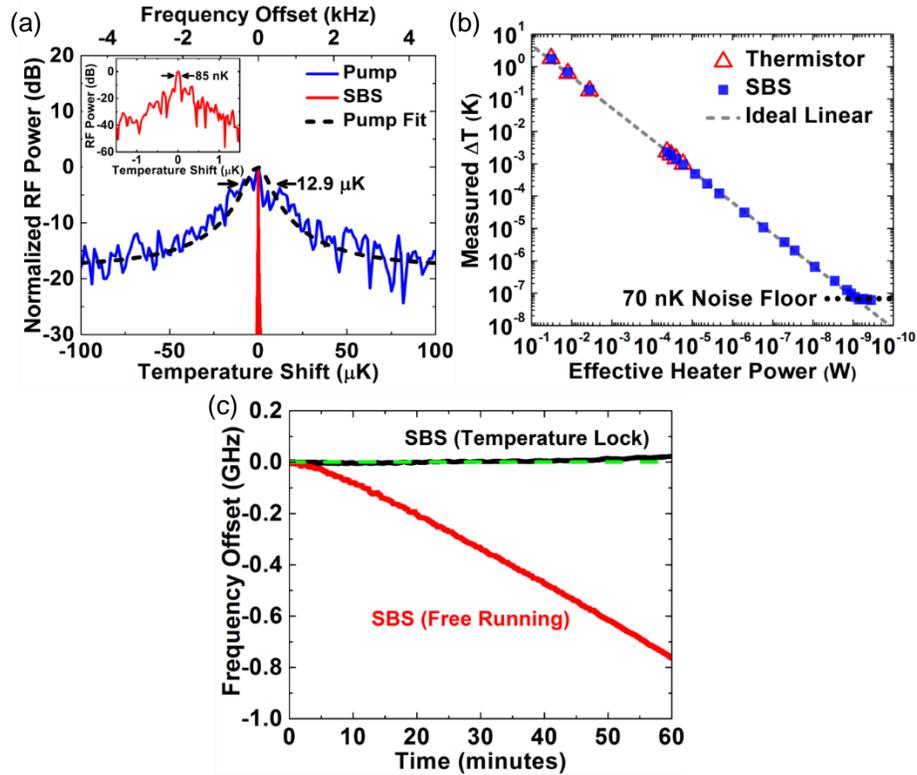

**Fig. 5.** SBS laser temperature stabilization. (a) Sensitivity of the locked pump (blue) and SBS laser (red) for the measurement of temperature. A fit to the pump laser (black dashed line) indicates a resolution of 12.9 µK, while the SBS laser line is a single narrow peak on the same scale. The inset shows a zoomed in plot of the SBS line indicating a temperature resolution of 85 nK. (b) Measurements of the temperature shift induced by a controlled variation in heater power. The thermistor (red open triangles) and SBS laser (blue squares) measurements were independently calibrated and agree with a linear dependence of temperature on heater power. (c) SBS laser frequency drift for free-running (red) and temperature stabilized (black) cases. When locked, the SBS drift is nearly unnoticeable compared to the ideal zero drift case (dashed green line).

To verify the sensitivity of our dual-mode SBS laser for temperature measurement, we directly induce a temperature change in our system by applying a known voltage shift to a resistive heater. The voltage shift is controlled through a tunable voltage source having 1 mV resolution, which is further attenuated through a variable attenuator having 1 dB step-size tunable attenuation. The shift in heater power is calculated from the change in voltage across an 80 Ω resistor, which then directly and linearly corresponds to a measured temperature shift. Figure 5(b) shows the induced temperature shift as measured by a calibrated thermistor and also independently measured by calibrating the frequency shift of our dual-mode SBS laser.

The three measurements at the top where the temperature shift is > 100 mK are performed with an applied DC variation in the heater power, since the room temperature fluctuations are not yet large enough to dominate over the applied shift in temperature. The rest of the measurements are performed using a 1 Hz applied sinusoidal heater power modulation where the strength of the induced 1 Hz temperature shift is recorded as a function of the modulation drive. We find excellent agreement between the thermistor and the SBS temperature sensor for larger temperature variations. Furthermore, by precisely controlling the attenuation of the applied voltage to even smaller values, we verify the temperature measurement to follow a linear dependence with heater power, as ideally expected. The last few measurement points confirm our ability to sense temperature change below 100 nK.

The exceptional temperature resolution that the SBS laser offers is ideally suited as a means to correct for the laser's own temperature variations over time. Given a detection sensitivity of 85 nK and also the ability to perform ideal temperature servoing, the frequency drift of the SBS laser is expected to reside within a bound of 140 Hz. In contrast to previous dual-mode demonstrations that stabilize the resonator temperature through varying optical intensity, we instead use direct resistive heating as our feedback mechanism to induce temperature change. As seen in Fig. 2(a), this distinction is critical in preventing a substantial Kerr shift from building up as the injected optical power changes, which would lead to a residual drift in the laser's frequency even as the temperature appears to be stabilized. Figure 5(b) shows the frequency excursion corresponding to one of the two SBS polarizations measured with and without temperature stabilization over a duration of an hour. Since both available pump lasers are consumed to measure the temperature of a single cavity, the absolute frequency of the SBS laser was instead measured using a wavelength meter exhibiting a few MHz resolution. In order to ensure the SBS laser frequency drift is large enough to be resolved, the temperature of the room was intentionally ramped to raise the room's temperature ~100× faster than typical laboratory conditions. From Fig. 5(b), the free-running SBS shifts by 760 MHz over an hour, while the temperature stabilized SBS exhibits no visible drift compared to the dashed-green line guide until the last 15 minutes of measurement. The stabilized SBS ends at a frequency 22 MHz shifted to the blue when our temperature feedback exits its intended range of operation. Taking a worst-case drift of 22 MHz per hour, the SBS laser frequency shifts by 6.1 kHz per second, which after accounting for the intentional ~100× faster applied temperature ramp, equates to ~60 Hz per second.

## 3. Summary

We demonstrated the importance of increasing the mode volume of a high-Q resonator for the purposes of increasing a resonator's resilience to temperature fluctuations. This technique combined with the ability to generate lasing from the SBS nonlinearity of an otherwise passive resonator allowed for the realization of a SBS laser that reaches a linewidth of 20 Hz. Owing to its inherent temperature stability, the SBS laser noise at low Fourier frequencies improves over the current state-of-the-art microresonator lasers [16, 20] by an order of magnitude. In order to prevent the long-term drift of the resonator's frequency, we applied the exquisitely narrow SBS line as an ultra-precise sensor to detect and correct for minute temperature shifts as small as 85 nK. Here, the SBS laser improved our ability to resolve temperature change by over two orders of magnitude compared to an otherwise identical dual-mode sensor employing a 270-Hz linewidth pump laser. These advances all serve to highlight the promise of the SBS laser as the next-generation precision source capable of replacing the ULE cavity laser for a growing body of applications that require an ultra-narrow linewidth and yet portable laser. With further development, we expect the SBS laser to achieve even narrower linewidths of 1 Hz and below by overcoming its current limitations of common-mode technical noise. In addition, although our present SBS laser system comprises discrete fiber-optic components, the recent innovations made in realizing high-Q microresonators on chip offer new and exciting possibilities for integrating the entirety of the SBS laser into a single monolithic package.

**Methods**

*A. Resonator Frequency Shift with Temperature*

In any resonator system, shifts in the resonance frequency ($v$) are related to refractive index ($n$) shifts and length ($L$) shifts by

$$\frac{dv}{v} \approx -\frac{dn}{n} - \frac{dL}{L} \tag{M1}$$

Both the refractive index and length are temperature dependent, which is made explicit in Eq. (M1) by including effects of thermoexpansion ($\alpha$) and thermally-induced changes in refractive index ($dn/dT$)

$$\frac{dv}{v} \approx -\frac{1}{n}\frac{dn}{dT}\Delta T - \alpha \Delta T \tag{M2}$$

For the material of silica glass, $\alpha = 0.51 \times 10^{-6}$ 1/K and $dn/dT = 11.6 \times 10^{-6}$ 1/K, which combined with $v = 193.6 \times 10^{12}$ Hz and n = 1.45 yields a 1.65 GHz shift in resonance frequency for a 1 °C shift in temperature.

**B. SBS Laser Thermal Response Measurement**

To measure the SBS laser's thermal response, we PDH lock the pump laser to the cavity resonance. We then apply an optical power modulation to the input pump light (via a semiconductor optical amplifier) or a temperature modulation to the resonator (via a resistive heater) and measure the resulting frequency shift response of the SBS laser. This experiment thus measures the coupling of the pump intensity or the resonator temperature to the SBS resonance's frequency shift, which we plot as a function of modulation frequency. These measurements were also performed for the locked pump laser's frequency shift and at low pump powers far below the Brillouin threshold with no observable difference in the response shape.

**C. Laser Frequency Noise Measurement**

We measured the frequency noise of the pump and PDH-locked pump lasers by using an unbalanced delay-line Mach-Zehnder with 250 m additional delay placed in one modulator arm. The frequency noise converts to voltage noise through the interferometer, which is measured on an electrical spectrum analyzer. The SBS laser's frequency noise was measured by direct heterodyning two independent SBS lasers and then sending the resulting microwave beat through a microwave frequency-to-voltage converter and into an electrical spectrum analyzer. For the common-cavity SBS configuration, the noise was below that of the frequency-to-voltage converter. We instead multiplied the microwave beat of the common-cavity SBS by a factor of 5 in order to increase the frequency noise above the floor of the converter.

### D. Temperature Sensitivity Measurements

For applied temperature variations above 100 mK, the responses of the thermistor and SBS laser are each independently measured and calibrated at DC. A controlled voltage step is applied to a resistive heater, and the initial and final thermistor readings are recorded to determine the resulting temperature change. Similarly, the same applied temperature shift is simultaneously measured by the dual-mode SBS laser, and the initial and final frequency beat between the modes are recorded. From the measured change in beat frequency, the slope of Fig. 4(c) is then applied to convert this frequency to an equivalent frequency shift of a single SBS laser line. The ratio of this frequency shift to the known 1.65 GHz / 1 °C yields the independently calibrated temperature as measured by the SBS laser.

For applied temperature shifts below 1 mK, the background temperature fluctuations of the environment dominate and prevent the ability to accurately measure small controlled changes in temperature. For these cases, the temperature sensitivity limit of the SBS laser is instead determined by using a lock-in approach where the heater power drive is modulated at a frequency of 1 Hz. The resulting 1 Hz temperature modulation becomes imprinted on the dual-mode SBS beat, which is then converted to a voltage signal via a frequency-to-voltage converter to be processed on a spectrum analyzer. The heater drive is successively lowered until the measurement reaches the SBS laser noise floor (see supplementary material).

### E. Temperature Stabilization of SBS Laser

We utilized both passive and active temperature stabilization to reduce the drift of our SBS laser system. Our passive temperature stabilization consisted of a plexiglass enclosure which isolates our SBS laser from environmental fluctuations. For active temperature control, we stabilized the SBS laser's temperature by first detecting the frequency of the dual-polarization SBS beat using a frequency counter. This measurement was then monitored on a computer, which used a software PID feedback loop to drive the voltage sent to a resistive heater for control of the resonator's temperature. A thermistor was also utilized to independently monitor the resonator temperature. The loop time constant was set to 0.5 s to prevent oscillation instability and to allow for SBS drift correction for Fourier frequencies below 1 Hz. This slow loop time constant also enables the SBS laser to maintain its excellent noise characteristics above 1 Hz frequency.

## 4. Acknowledgments

We thank Prof. Rajeev J. Ram at the Massachusetts Institute of Technology and also Dr. Sumanth Kaushik, Dr. Matthew Stowe, Dr. David Caplan, and Dr. Robert McConnell at MIT Lincoln Laboratory for helpful discussions. The opinions, interpretations, conclusions, and recommendations are those of the authors and are not necessarily endorsed by the United States Government.